\documentclass[espf]{aastex}
\usepackage{emulateapj5}

\slugcomment{Printed \today}

\shorttitle{The Remarkable Mid-Infrared Jet of G35.20-0.74}
\shortauthors{De Buizer}

\begin{document}

\title{The Remarkable Mid-Infrared Jet of Massive Young Stellar Object G35.20-0.74}

\author{James M. De Buizer}
\affil{Gemini Observatory, Casilla 603, La Serena, Chile;
jdebuizer@gemini.edu  }

\begin{abstract}

The young massive stellar object G35.20-0.74 was observed in the
mid-infrared using T-ReCS on Gemini South. Previous observations
have shown that the near infrared emission has a fan-like morphology
that is consistent with emission from the northern lobe of a bipolar
radio jet known to be associated with this source. Mid-infrared
observations presented in this paper show a monopolar jet-like
morphology as well, and it is argued that the mid-infrared emission observed is
dominated by thermal continuum emission from dust. The mid-infrared
emission nearest the central stellar source is believed to
be directly heated dust on the walls of the outflow cavity. The
hydroxyl, water, and methanol masers associated with G35.20-0.74 are
spatially located along these mid-infrared cavity walls.
Narrow jet or outflow cavities such as this may also be the
locations of the linear distribution of methanol masers that are
found associated with massive young stellar objects. The fact that
G35.20-0.74 has mid-infrared emission that is dominated by the
outflow, rather than disk emission, is a caution to those that
consider mid-infrared emission from young stellar objects as only
coming from circumstellar disks.

\end{abstract}


\section{Introduction}

While molecular outflows are an apparently ubiquitous phenomena in
regions of high-mass star formation (Shepherd and Churchwell 1996;
Zhang et al. 2001; Beuther et al. 2002), clear examples of
individual young massive stars with well-defined bipolar jets are
relatively few. One such example is G35.20-0.74, a massive star
formation region containing an early B-type star surrounded by an
ultracompact HII region (UCHII region) with a bipolar jet-like radio
structure. Water, OH, and methanol masers (Forster \& Caswell 1989,
Hutawaraforn \& Cohen 1999) are all associated with this massive
young stellar object, though the nature of their relationship with
respect to G35.20-0.74 is still unclear.

This jet and outflow have recently been observed in the
near-infrared (NIR) by Fuller, Zijlstra, \& Williams (2001) and in the
mid-infrared (MIR) at low spatial resolution ($\sim$1$\arcsec$) by De
Buizer et al. (2005). The nature and structure of the thermal
infrared emission seen from this outflow in these two studies has
prompted this high spatial resolution ($\sim$0.35$\arcsec$ at
11.7$\micron$) MIR follow-up study. In this Letter, I will
explore the characteristics of the MIR emission in
G35.20-0.74 and relationships between the detailed morphologies of
the infrared emission, masers, and radio continuum emission.

\section{Observations}

Observations of G35.20-0.74 were carried out at Gemini South on the
night of 2005 July 12 through patchy clouds. Imaging was performed
with the Thermal-Region Camera and Spectrograph (T-ReCS) using the
\emph{Si-5} ($\lambda$$_c$=11.7$\micron$,
$\Delta\lambda$=1.1$\micron$) filter and \emph{Qa} filter
$\lambda$$_c$=18.3$\micron$, $\Delta\lambda$=1.6$\micron$). T-ReCS
utilizes a Raytheon 320$\times$240 pixel Si:As IBC array which is
optimized for use in the 7--26 $\mu$m wavelength range. The pixel
scale is 0.089$\arcsec$/pixel, yielding a field of view of
28$\farcs$8$\times$21$\farcs$6. Co-added frames were
saved every 10 sec, and the telescope was nodded every 30 sec. The
co-added frames were examined individually during the data reduction
process and those plagued by clouds (i.e., showing high and/or
variable background or decreased source flux) were discarded. In
both filters the northern and southern radio jet regions were imaged, yielding final
mosaicked images with effective field of views of approximately
28$\arcsec$$\times$35$\arcsec$. The 8$\arcsec$ overlapping region was used to
register the northern and sourthern images of each mosaic.
However, the individual images were cropped before they were
mosaicked so the final mosaic has the same signal to noise across
the whole image. The final \emph{effective} exposure times for the mosaics
presented here are therefore 140 sec at 11.7$\micron$ and 180 sec at
18.3$\micron$.

Flux calibration was achieved by observing the MIR standard
star HD 169916 ($\lambda$ Sgr) at similar airmasses to the G35.20-0.74 observations.
The assumed flux densities for HD 169916 were taken to be 22.29 Jy
at 11.7$\micron$ and 9.24 Jy at 18.3$\micron$. These assumed
standard star flux densities were found by convolving the spectral
irradiance templates of the stars from Cohen et al. (1999) with the
given T-ReCS filter transmission profile. Derived flux densities for
the entire region of G35.20-0.74 are measured to be 3.06$\pm$0.09 Jy
at 11.7 $\micron$ and 46.87$\pm$3.66 Jy at 18.3 $\micron$ using a
square aperture of 6$\farcs$9$\times$13$\farcs$5. These flux
densities are quoted with their 1-$\sigma$ total error, which is a
quadrature addition of the statistical variation from the aperture
photometry (due to the standard deviation of the background array
noise) and the flux calibration error. The flux calibration error
was found from the standard deviation of the variation of the
standard star flux density in each co-added nod position. Since the
science and standard star images were cleaned of any effects from
clouds, the quoted errors are thought to be robust, however the
errors were calculated using all available data. Comparisons to the
lower angular resolution observations by De Buizer et al. (2005)
show the values derived here to be consistent with those derived in
that previous work.

\section{Discussion}

The mosaicked images at 11.7 and 18.3 $\micron$ are presented in
Figure 1. These images are cropped to show only the parts of the
field that have MIR sources. No significant MIR
emission was detected outside these cropped areas at a 3-$\sigma$
upper limit of 41 mJy$\cdot$arcsec$^{-2}$ at 11.7 $\micron$ and 283
mJy$\cdot$arcsec$^{-2}$ at 18.3 $\micron$. Sources 1, 2, and 10 are
seen at 11.7 $\micron$ but not at 18.3 $\micron$, and source 3 is
seen at 18.3 $\micron$, but not 11.7 $\micron$. Source 4 is
marginally detected at 18.3 $\micron$. The remaining sources are
detected at both wavelengths and are mostly knots of emission
associated with the MIR monopolar jet of G35.20-0.74. The
origin of Figure 1 is the expected location of the outflow source
itself. This source is a B2.6 star (as derived from the 8.5GHz flux
density of Gibb et al. 2003 and using the method described in De
Buizer et al. 2005) that can be seen as a UC HII region in the radio
and has been dubbed G35.2N.

\subsection{Relations to radio continuum and NIR emission}

The MIR images were registered with respect to the
NIR K and L$^\prime$ images of Fuller et al. (2001). Very
accurate relative astrometry ($<$0.15$\arcsec$) was achieved because
of the presence of three compact MIR sources (1, 2, and 4)
that are also present in the K and/or L$^\prime$ images. The
absolute astrometry of the NIR images (and, consequently,
the MIR images) comes from matching up NIR point
sources with their optical counterparts found in the USNO B1.0
astrometric catalog. The estimates of the 1-$\sigma$ absolute
uncertainty in these coordinates are 0.3$\arcsec$ for Right
Ascension and 0.1$\arcsec$ for Declination.

Figure 2a shows the K emission (thick white contours) overlaid on
the 11.7 $\micron$ image, and Figure 2b shows the L$^\prime$
emission (thin white contours) overlaid on the 18.3 $\micron$ image.
The L$^\prime$ emission from the jet looks very similar to what is
seen in the MIR. The K emission appears to be dominated
more by the material in the north, further along the outflow axis,
with very little emission down near the outflow source itself. The
convex structure seen in the NIR breaks up into separate
MIR components (knots 5 and 8), and therefore is probably
not a bow-shock as implied by Fuller et al. (2001).

The L$^\prime$ images of Fuller et al. (2001) show what they claim
is weak NIR emission from the southern jet of G35.20-0.74
(see Figure 2b). Interestingly, this emission is extremely weak at
K, bright at L$^\prime$, not detected at 11.7 $\micron$, but present
at 18.3 $\micron$ (Source 3 in Figure 1).

The MIR images were also registered with respect to the
high-resolution 8.5 GHz radio continuum images of Gibb et al. (2003)
in Figure 2a (thin white contours), and with the low-resolution 15
GHz radio continuum image of Heaton \& Little (1988) in Figure 2b
(thick white contours). The 1-$\sigma$ relative astrometric error
between the MIR and radio continuum images is estimated to
be 0.34$\arcsec$ in Right Ascension and 0.18$\arcsec$ in
Declination. The MIR and NIR images and contours
shown in Figure 2 have been shifted +0$\fs$023 (+0.35$\arcsec$) in
Right Ascension to place G35.N on the infrared outflow axis (this is
approximately the estimated 1-$\sigma$ astrometric uncertainty).

In Figure 2b it can be seen that the overall extent of the northern
radio lobe is comparable to the MIR emission. There is also
considerable MIR emission coming from the central radio
continuum emitting region near the outflow source, however there is
no MIR emission from the southern radio peaks. G35.2N is
also the location of one of the two millimeter peaks (Gibb et al.
2003) in this region (crosses in Figure 2b).

The two northernmost radio knots lie close to, but are not exactly
coincident with MIR sources 6 and 7. For these knots, the
radio and the MIR may be tracing slightly different
emitting regions within the knots themselves.

\subsection{Nature of the mid-infrared emission}

MIR emission from outflows has been detected previously (e.g., Noriega-Crespo
2004), however these outflows have been claimed to be dominated by
shock lines of H$_2$ contained within the filters used. For the
observations presented here there are no H$_2$ lines within the
bandpass of either the 11.3 or 18.3 $\micron$ filters. There is a possibility that there may be some contribution to the emission at 11.7 $\micron$ because of PAH emission, however this is not a concern at 18.3 $\micron$. The steeply rising spectral slope from L$^\prime$ to 18.3 $\micron$ of the narrow, elongated infrared
emission coincident with and immediately north of the position of G35.2N demonstrates that the infrared emission here is dominated by longer wavelength continuum emission. Therefore the nature of the
infrared emission is concluded to be dominantly continuum dust emission from the outflow cavity walls. This
cavity was created by the molecular outflow punching a hole in the
dense molecular material surrounding young stellar source at the
center of G35.20-0.74. The central source is mostly likely directly
heating the walls of this cavity. The northern lobe of the outflow
was found to be slightly blueshifted towards Earth (i.e., in CO by
Gibb et al. 2003; in CI by Little, Kelly, \& Murphy 1998). Given
this fortuitous geometry, we can see directly into the outflow
cavity itself due to the outflow clearing away the material along
our line of sight.

The sources further north of G35.20-0.74, namely sources 5-9, are
expected to be knots of dust either in the outflow itself of clumps
of pre-existing material that are being impinged upon by the
outflow. Source 6 lies 19200 AU from G35.2N and is still at an estimated dust color
temperature of 112 K. This is based on the 11.7 and 18.3 $\micron$ flux densities of this source and neglecting the possible effects of silicate absorption (see De Buizer et al. 2005 for method and limitations). What is heating the dust this far out? Smaller
dust grains can be heated out to farther distances than large dust
grains. The typical size range of interstellar grains is believed to
be 0.003-10 $\micron$, and typical grain compositions include smooth
astronomical silicates, graphite, and silicon carbide (Laor \&
Draine 1993, Draine \& Lee 1984). For the following I will use the
equation for dust temperature given by Sellgren, Werner, \&
Dinerstein (1983) and the ultraviolet and infrared emissivities of
Draine \& Lee (1984). Assuming the dust is made up of smooth
astronomical silicates, dust with a lower limit size of 0.003
$\micron$ can be heated to 112K only out to $\sim$16000 AU by a B2.6
star. If the dust is made of graphite, one can heat out to the
distance of source 6 with grains having a typical size of 0.005
$\micron$, still near the lower limit size. However, if silicon
carbide is the assumed composition of the dust, then one can get
heating out much farther than source 6, namely $\sim$52000 AU at the
0.003 $\micron$ lower limit size. There is a possibility of some
contribution from shock heating, although Fuller et al. 2001 claim
no detection of shock-excited H$_2$ in the region. Beaming of the
MIR emission along the outflow axis, rather than the
isotropic emission assumed in the above calculations, could also
help in heating grains farther out.  Interestingly, the MIR
luminosity derived from the dust color temperature gives an estimated value of 1.6$\times$10$^3$
L$_{\sun}$. Assuming the MIR luminosity is all the
luminosity of the source (an obvious underestimate) and calculating
a spectral type from that bolometric luminosity using the method of
De Buizer et al. (2005) gives a value of $\sim$B3, consistent with the
radio derived spectral type. In summary, all of the dust, even as far
out as source 6, \emph{can} indeed be heated directly by G35.2N,
depending on dust composition and size (as well as beaming), though
we cannot rule out contributions from other possible heating mechanisms.

As discussed in the previous section, MIR source 3,
coincident with NIR emission from the presumed infrared
southern counterjet, does not have a smoothly increasing spectral
slope typical of dust continuum emission, but instead is only
present at L$^\prime$ and 18.3 $\micron$. This implies that the
emission in this southern source is dominated by line emission of
some kind. The usual suspects are: 1) H$_2$ emission from shocks,
however, Fuller et al. (2001) claim no detection of H$_2$ in the
region; 2) PAH emission from the photo-dissociation region of the
outflow interface with the molecular cloud, however the L$^\prime$
and 18.3 $\micron$ filters do not encompass any PAH features; 3)
[FeII] emission from shocks. This last one may be a possibility
since [FeII] lines are found in both filter bandpasses, but
spectroscopy will be needed to know for sure.

\subsection{Masers}

The positions of the OH masers from Hutawarakorn \& Cohen (1999) and
water masers from Forster \& Caswell (1989) are shown in Figure 2c
(asterisks and X symbols, respectively). The positions of these
maser sources with respect to the radio continuum emission were
taken from Gibb et al. (2003). There are also methanol masers
plotted in Figure 2c (large crosses), that were found by A.G. Gibb
(private communication), and have positions known to 0.2$\arcsec$
with respect to the radio continuum.

Interestingly, the combined distribution of the masers is in a
V-shape, with its apex near the G35.2N source itself. From their
positions with respect to the near and MIR emission, the
masers appear to be tracing the outflow cavity walls. In the case of
the OH and water masers, they may be excited to emit here by the
slower oblique shocks created by the outflow on the cavity walls.
Methanol masers on the other hand are believed to be excited by
MIR radiation (i.e., Sobolev \& Deguchi 1994; Sobolev,
Cragg, \& Godfrey 1997). The copious amount of MIR emission
associated with this outflow source implies that at least the
pumping mechanism needed for the generation of methanol masers
exists on such outflow cavity walls.

Given the narrow opening angle of this outflow cavity as seen in the
MIR and the fact that methanol masers are often found in
linear distributions (e.g. Norris et al 1993), it is possible that
methanol masers in general are associated with the cavities walls of
outflowing young massive stars. Previous observations by De Buizer
(2003) found the majority of linear methanol maser distributions in
that sample were found at position angles similar to H$_2$ emission
expected to trace outflows from the host YSOs. Furthermore, other
sources like NGC 7538 IRS 1 (De Buizer \& Minier 2005) have been
observed in the MIR to have outflow-like MIR
emission at the same position angle as other outflow indicators (CO
in the case of NGC 7538 IRS1), as well as having a linear methanol
maser distribution at a similar position angle as well. All of this
may point to a scenario where the linearly distributed methanol
masers may be associated more generally with outflows of massive
YSOs, and not with circumstellar disks as was previously thought
(i.e. Norris et al 1993).

\subsection{Implications for mid-infrared bright YSOs}

There is a tendency to think that if MIR emission is
detected in a young stellar object (especially if it appears
elongated in its morphology) that it is emission from a
circumstellar disk (i.e. Lada \& Lada 2003). Recent observations and theories (e.g.,
Miroshnichenko et al. 1999) have shown that, at least for the more
massive YSOs, the dominant source of MIR emission
comes from an accretion envelope. These accretion envelopes may be
elongated as well (De Buizer, Osorio, \& Calvet 2005). The
observations presented here show that we must take in to account
another source of elongated MIR emission. On a 3-4m
telescope (De Buizer et al. 2005), or with the resolution of the
\emph{Spitzer Space Telescope}, the MIR emission from this
G35.20-0.74 could be misinterpreted as a circumstellar disk. Only
recently with the increase of MIR instruments on 8-10m
telescopes have we begun to resolve source like G35.20-0.74 with
such detail as to discern their true MIR emission as being
outflow-related (e.g., NGC7538 IRS 1; De Buizer \& Minier 2005).

Therefore one must be cautious when trying to infer disk properties
from unresolved or partially resolved sources since the outflow
cavities of YSOs can be the dominant source of MIR
emission. On the other hand, the mere presence of such outflow
cavities with small opening angles implies the presence of
collimating accretion disks at the bases of the outflows. So while
MIR emission may not always be a \emph{direct} tracer of
circumstellar disks and their properties, the presence of such
emission in the near circumstellar environment of YSOs may still
indirectly indicate the presence of a disk.

\acknowledgments Based on observations obtained at the Gemini Observatory,
which is operated by AURA, Inc., under a cooperative agreement with the NSF on behalf of
the Gemini partnership: the NSF (U.S.), the PPARC (U.K.), the NRCC (Canada), CONICYT (Chile), the ARC (Australia), CNPq (Brazil) and CONICET (Argentina). I would also like to thank James Radomski for his
illuminating discussions which helped in the writing of this article.

\clearpage

\begin{figure}
\epsscale{0.95}
\plotone{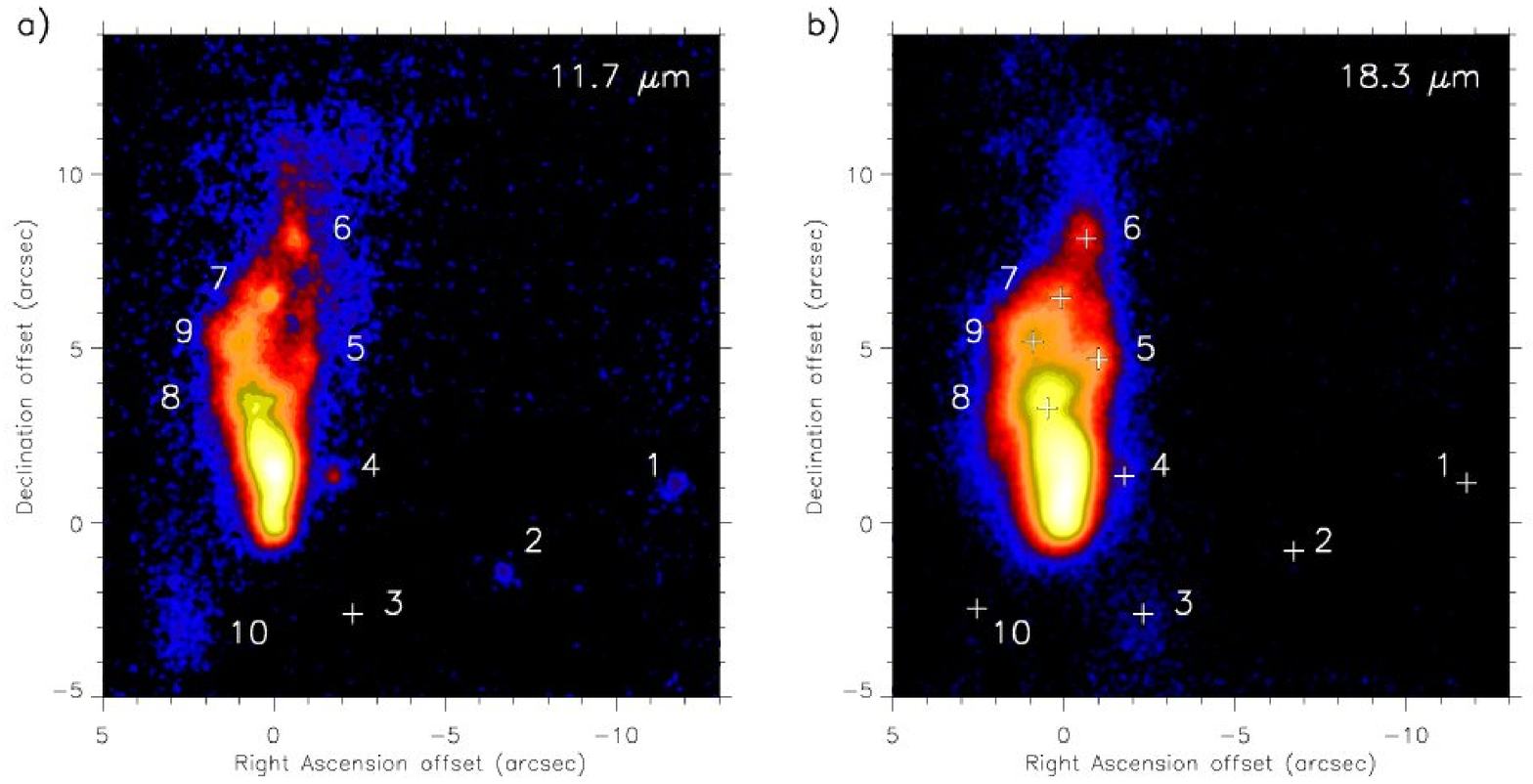}
\caption{The region of
G35.20-0.74 in false-color as seen at a) 11.7 $\micron$ and b) 18.3
$\micron$ with T-ReCS. Crosses in b) show the locations of
individual MIR sources at 11.7 $\micron$ and are numbered
by increasing right ascension. The cross in a) shows the 18.3
$\micron$ location of source 3 which is not seen at 11.7 $\micron$.}
\label{fig1}
\end{figure}

\clearpage

\begin{figure}
\epsscale{1.0}
\plotone{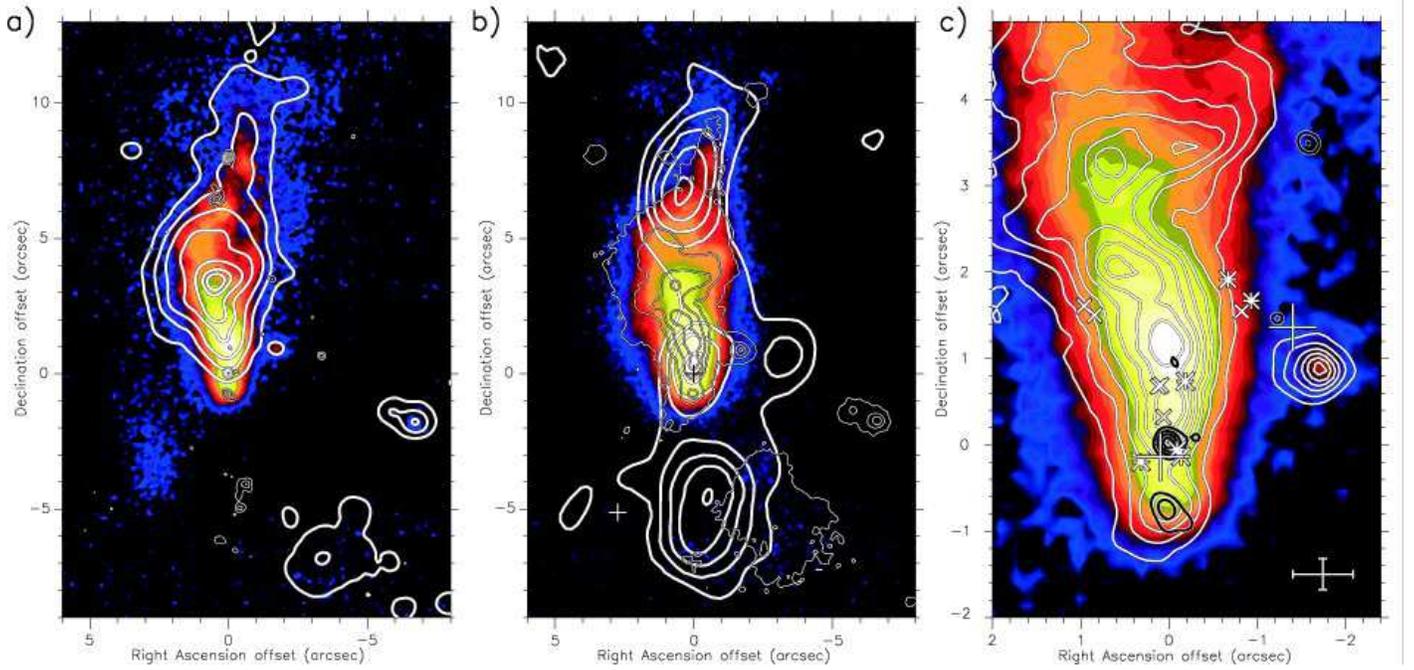}
\caption{The G35.20-0.74 jet
as seen at different wavelengths. Panel a) shows the 11.7 $\micron$
image in false-color overlaid by K-band emission of Fuller et al.
(2001) as thick white contours, and the 8.5 GHz high resolution
radio continuum emission of Gibb et al. (2003) as thin white
contours. Panel b) shows the 18.3 $\micron$ image in false color
overlaid by the low resolution 15 GHz radio continuum image of
Heaton \& Little (1988) as thick white contours, and L$^\prime$
image of Fuller et al. (2001) as the thin white contours. Panel c)
is a zoom in on the central region of the 11.7 $\micron$ image in
false-color, the L$^\prime$ contours in white and the high
resolution radio continuum contours in black. OH maser of
Hutawaraforn \& Cohen (1999) are shown as asterisks, water masers of
Forster \& Caswell (1989) are shown as X symbols, and methanol
masers of A. G. Gibb (priv. comm.) shown as large crosses. The cross
in the lower right of panel c) shows the $\pm$1-$\sigma$ relative
astrometric uncertainty between the radio continuum and
NIR.} \label{fig2}
\end{figure}


\begin{thebibliography}{}

\bibitem[Beuther et al.(2002)]{BSSMW02}Beuther, H., Schilke, P., Sridharan,
T.~K., Menten, K.~M., Walmsley, C.~M., \& Wyrowski, F. 2002, \aap,
383, 892

\bibitem[Cohen et al.(1999)]{1999AJ....117.1864C} Cohen, M., Walker, R.~G.,
Carter, B., Hammersley, P., Kidger, M., \& Noguchi, K.\ 1999, \aj,
117, 1864

\bibitem[De Buizer et al.(2005)]{2005ApJS..156..179D} De Buizer, J.~M.,
Radomski, J.~T., Telesco, C.~M., \& Pi{\~n}a, R.~K.\ 2005, \apjs,
156, 179

\bibitem[De Buizer \& Minier(2005)]{2005ApJ...628L.151D} De Buizer, J.~M.,
\& Minier, V.\ 2005, \apjl, 628, L151

\bibitem[De Buizer(2003)]{2003MNRAS.341..277D} De Buizer, J.~M.\ 2003,
\mnras, 341, 277

\bibitem[Forster \& Caswell(1989)]{1989A&A...213..339F} Forster, J.~R., \&
Caswell, J.~L.\ 1989, \aap, 213, 339

\bibitem[Fuller et al.(2001)]{2001ApJ...555L.125F} Fuller, G.~A., Zijlstra,
A.~A., \& Williams, S.~J.\ 2001, \apjl, 555, L125

\bibitem[Gibb et al.(2003)]{2003MNRAS.339.1011G} Gibb, A.~G., Hoare, M.~G.,
Little, L.~T., \& Wright, M.~C.~H.\ 2003, \mnras, 339, 1011

\bibitem[Heaton \& Little(1988)]{1988A&A...195..193H} Heaton, B.~D., \&
Little, L.~T.\ 1988, \aap, 195, 193

\bibitem[Hutawarakorn \& Cohen(1999)]{1999MNRAS.303..845H} Hutawarakorn,
B., \& Cohen, R.~J.\ 1999, \mnras, 303, 845

\bibitem[Lada \& Lada(2003)]{2003ARA&A..41...57L} Lada, C.~J., \& Lada,
E.~A.\ 2003, \araa, 41, 57

\bibitem[Little et al.(1998)]{1998MNRAS.294..105L} Little, L.~T., Kelly,
M.~L., \& Murphy, B.~T.\ 1998, \mnras, 294, 105

\bibitem[Miroshnichenko et al.(1999)]{1999ApJ...520L.115M} Miroshnichenko,
A., Ivezi{\'c} , {\v Z}., Vinkovi{\'c} , D., \& Elitzur, M.\ 1999,
\apjl, 520, L115

\bibitem[Noriega-Crespo et al.(2004)]{2004ApJS..154..352N} Noriega-Crespo,
A., et al.\ 2004, \apjs, 154, 352

\bibitem[Norris et al.(1993)]{1993ApJ...412..222N} Norris, R.~P., Whiteoak,
J.~B., Caswell, J.~L., Wieringa, M.~H., \& Gough, R.~G.\ 1993, \apj,
412, 222

\bibitem[Sellgren et al.(1983)]{1983ApJ...271L..13S} Sellgren, K., Werner,
M.~W., \& Dinerstein, H.~L.\ 1983, \apjl, 271, L13

\bibitem[Shepherd \& Churchwell(1996)]{SC96}Shepherd, D.~S., \& Churchwell,
E. 1996, \apj, 457, 267

\bibitem[Sobolev \& Deguchi(1994)]{1994A&A...291..569S} Sobolev, A.~M., \&
Deguchi, S.\ 1994, \aap, 291, 569

\bibitem[Sobolev et al.(1997)]{1997A&A...324..211S} Sobolev, A.~M., Cragg,
D.~M., \& Godfrey, P.~D.\ 1997, \aap, 324, 211

\bibitem[Zhang et al.(2001)]{Z01}Zhang, Q., Hunter T.~R., Brand, J.,
Sridharan, T.~K., Molinari, S., Kramer, M.~A., \& Cesaroni, R. 2001,
\apj, 552, 167


\end{thebibliography}
\end{document}